\documentstyle[prl,aps,epsf,epsfig]{revtex}
\begin{document}

\title{Bohmian trajectories for photons}

\author{Partha Ghose$^1$, A. S. Majumdar$^1$, S. Guha$^2$ and J. Sau$^2$}

\address{$^1$S. N. Bose National Centre for Basic Sciences,
Block JD, Sector III, Salt Lake, Kolkata 700 098, India}

\address{$^2$Department of Electrical Engineering, Indian Institute of Technology, Kanpur 
208 016, India}

\maketitle

\newcommand{\be}{\begin{equation}}
\newcommand{\ee}{\end{equation}}
\newcommand{\ben}{\begin{eqnarray}}
\newcommand{\een}{\end{eqnarray}}

\begin{abstract}
The first examples of Bohmian trajectories for photons have been worked
out for simple situations, using the Kemmer-Duffin-Harishchandra formalism.

PACS NO(s): 03.65.Bz
\end{abstract}

\section{Introduction}

It is generally believed that only massive fermions have Bohmian trajectories 
but bosons do not. This is usually attributed to the impossibility of 
constructing a relativistic quantum mechanics of bosons with a conserved 
four-vector probability current density with a positive definite time
component. However,
it has now been shown \cite{Ghose} that a consistent relativistic quantum
mechanics of spin 0 and spin 1 bosons can be developed using the
Kemmer equation \cite{Kemmer}

\be
(\,i\,\hbar\,\beta_\mu\,\partial^\mu + m_0\,c\,)\,\psi = 0\,
\label{eq:1}
\ee
where the matrices $\beta$ satisfy the algebra

\be
\beta_{\mu}\,\beta_{\nu}\,\beta_{\lambda} + \beta_{\lambda}\,\beta_{\nu}\,
\beta{\mu} = \beta_{\mu}\,g_{\nu \lambda} + \beta_{\lambda}\,g_{\nu \mu}\,.
\label{eq:2}
\ee
The $5\times 5$ dimensional representation of these matrices describes spin 0
bosons and the $10 \times 10$ dimensional representation describes spin 1
bosons. The fact that a conserved four-vector current with a positive
definite time component can be defined using this formalism can be seen as 
follows.  Multiplying (\ref{eq:1}) by $\beta_0$, one obtains the
Schr\"{o}dinger form of the equation

\be
i\,\hbar\,\frac{\partial \psi}{d t} = [\,- i\,\hbar\,c\, \tilde{\beta}_i\,
\partial_i - m_0\,c^2\,\beta_0\,]\,\psi
\label{eq:3}
\ee
where $\tilde{\beta}_i \equiv \beta_0\,\beta_i - \beta_i\,\beta_0$. Multiplying
(\ref{eq:1}) by $1- \beta_0^2$, one obtains the first class constraint

\be
i\,\hbar\,\beta_i\,\beta_0^2\,\partial_i\,\psi = -m_0\,c\,(\,1 - \beta_0^2\,)
\,\psi.
\label{eq:4}
\ee
It implies the four conditions $\vec{A}=\vec{\nabla}\times  \vec{B}$ and $\vec{\nabla}. \vec{E}=0$ 
in the spin-$1$ case. The reader is referred to Ref. \cite{Ghose} for further discussions regarding
the significance of this constraint.

If one multiplies equation (\ref{eq:3}) by $\psi^{\dagger}$ from the left,
its hermitian conjugate by $\psi$ from the right and adds the resultant
equations, one obtains the continuity equation

\be
\frac{\partial\,( \psi^{\dagger}\,\psi )}{\partial t} + \partial_i\,
\psi^{\dagger}\,\tilde{\beta}_i\,\psi = 0\,.
\label{eq:5}
\ee
This can be written in the form

\be
\partial^\mu\,s_{\mu} = 0
\ee
where $s_{\mu}=-\Theta_{\mu\nu}a^{\nu}$ (with $a^{\nu}a_{\nu}=1$ where $a^{\nu}$
is the unit four-velocity of the observer),  
$\Theta_{\mu \nu} = -m_0c^2\bar{\psi}(\beta_{\mu}\beta_{\nu}+\beta_{\nu}\beta_{\mu} - 
g_{\mu\nu})\psi$ 
is the symmetric energy-momentum tensor so that
$\Theta_{0 0} = - m_0\,c^2\,\psi^{\dagger}\,\psi < 0$. 
Notice that $s_{\mu}s^{\mu}=\Theta_{\mu\nu}\Theta^{\mu\nu} \ge 0$, so that
$s_{\mu}$ is time-like.
Thus, it is possible to  define
a wave function $\phi = \sqrt{m_0\,c^2/E}\,\psi$ (with $E =- \int\,\Theta_{0 0}
\,dV$ ) such that
$\phi^{\dagger}\,\phi$ is non-negative and normalized and can be
interpreted as a probability density. The conserved probability current
density is $s_\mu = - \Theta_{\mu 0}/E = (\,\phi^{\dagger}\,\phi,
- \phi^{\dagger}\,\tilde{\beta}_i\,\phi )$ \cite{Ghose}.

Notice that according to the equation of motion (\ref{eq:3}), the velocity
operator for massive bosons is $c\,\tilde{\beta}_i$, so that the Bohmian
3-velocity can be defined by

\be
v_i = \frac{d x_i}{d t} = \gamma^{- 1}\,u_i = c\,\frac{u_i}{u_0} = c\,\frac{s_i}{s_0}
= c\,\frac{\psi^{\dagger}\,\tilde{\beta}_i\,\psi}{\psi^{\dagger}\,\psi}\,.
\label{eq:6}
\ee
It follows from equation (3) that $c\tilde{\beta}_i$ is the velocity operator
whose eigenvalues are $\pm c$. Therefore, $v_{\mu}v^{\mu} =0$, and so the
Bohmian velocity is always timelike.
Integrating equation (7), one obtains a system of Bohmian trajectories $x_i(t)$ corresponding to different initial positions of the particle. In Bohmian mechanics one assumes that the initial distribution of the positions is given by $\vert \psi(0)\vert^2$. The continuity equation (\ref{eq:5}) then guarantees that the distribution will agree with quantum mechanics at all future times. The (Gibbs) ensemble averages of all dynamical variables in Bohmian mechanics will therefore always agree with the expectation values of the corresponding hemitian operators in quantum mechanics. 

The theory of massless spin 0 and spin 1 bosons cannot be obtained simply by
taking the limit $m_0$ going to zero. One has to start with the equation
\cite{HC}

\be
i\,\hbar\,\beta_\mu \partial^\mu\,\psi + m_0\,c\,\Gamma\,\psi = 0
\label{eq:8}
\ee
where $\Gamma$ is a matrix that satisfies the following conditions:

\ben
\Gamma^2 &=& \Gamma\,\\
\Gamma\,\beta_\mu + \beta_\mu\,\Gamma &=& \beta_\mu\,.
\label{eq:9}
\een
Multiplying (\ref{eq:8}) from the left by $1 - \Gamma$, one obtains

\be
\beta_\mu\,\partial^\mu\, (\,\Gamma\,\psi\,) = 0\,.
\label{eq:10}
\ee
Multiplying (\ref{eq:8}) from the left by $\partial_{\lambda}\,
\beta^{\lambda}\,\beta^{\nu}$, one also obtains

\be
\partial^{\lambda}\,\beta_{\lambda}\,\beta_\nu\,(\,\Gamma\,\psi\,) =
\partial_\nu\, (\,\Gamma\,\psi\,)\,.
\label{eq:11}
\ee
It follows from (\ref{eq:10}) and (\ref{eq:11}) that

\be
\Box\,\, (\,\Gamma\,\psi\,) = 0
\label{eq:12}
\ee
which shows that $\Gamma\,\psi$ describes massless bosons. The Schr\"{o}dinger
form of the equation

\be
i\,\hbar\,\frac{\partial\, (\,\Gamma\,\psi\,)}{d t} = - i\,\hbar\,c
\tilde{\beta}_i\,\partial_i\, (\Gamma\,\psi)
\label{eq:13}
\ee
and the associated first class constraint

\be
i\,\hbar\,\beta_i\,\beta_0^2\,\,\partial_i\,\psi
+ m_0\,c\,(\,1 - \beta_0^2\,)\,\Gamma\,\psi = 0
\label{eq:14}
\ee
follow by multiplying (\ref{eq:8}) by $\beta_0$ and $1 - \beta_0^2$
respectively. Equation (14) implies the Maxwell
equations ${\rm curl} \vec{E} = -(\mu/c)
\partial_t \vec{H}$ and ${\rm curl} \vec{H} = (\epsilon/c)
\partial_t \vec{E}$ if
\be
\Gamma \psi^T = (1/\sqrt{m_0 c^2}) \\ ( -D_x, -D_y, -D_z, B_x,
B_y, B_z, 0, 0, 0, 0)\nonumber\ee The constraint (15)
implies the relations ${\rm div} \vec{E} = 0$ and $\vec{B} = {\rm
curl}\vec{A}$. The symmetrical energy-momentum tensor is
\be
\Theta_{\mu \nu} = -\frac{ m_0 c^2}{2} \bar{\psi}(\beta_{\mu}
\beta_{\nu} + \beta_{\nu} \beta_{\mu} - g_{\mu \nu})\Gamma\psi\ee
and so the energy density

\ben {\cal{E}} = -\Theta_{0 0} = \frac{m_0 c^2}{2}
\psi^{\dagger}\,\Gamma \psi =  \frac{1}{2}[\vec{E}. \vec{E} +
\vec{B}. \vec{B} ]\een is positive definite.
The rest of the arguments are analogous to the massive case.

The Bohmian 3-velocity $v_i$ for massless bosons can be defined by

\be
v_i = c \frac{\psi^T \Gamma \tilde{\beta}_i \Gamma\psi}{\psi^T
\Gamma \psi}
\ee
Using arguments similar to the case of massive bosons, it is easy to see
that the Bohmian velocities for massless bosons are also timelike.
 Integrating equation (19) with different initial positions, one gets a system of Bohmian trajectories for the photon.
Neutral massless vector bosons are very special in quantum mechanics. Their
wave function is real, and so their charge current $j_\mu =
\phi^{T}\,\beta_\mu\,\phi$ vanishes. However,
their probability current density $s_\mu$ does not vanish.
Furthermore, $s_i$ turns out to be proportional to the Poynting vector, as it
should.

In this paper we compute Bohmian trajectories for photons
for certain simple but interesting cases. Integral curves of the Poynting vector for localized 
wave packets in classical electrodynamics were first plotted by Prosser\cite{Prosser}. 
They are lines of energy flow in classical electrodynamics and cannot be 
interpreted as particle trajectories. A particle trajectory interpretation
of these curves is possible  only within the context of  a proper
relativistic quantum mechanics of indivisible photons. This is what we have
done to calculate Bohmian velocities and
hence trajectories for photons, carrying the entire interpretational
package of Bohmian mechanics. It is only incidental that such trajectories
happen to coincide with the integral curves of the Poynting vector for
single photons. However, in the case of two photons, the Bohmian 
trajectories are computed from a two photon symmetrized wave function which
has no classical analogue. In this sense, 
the Bohmian trajectories calculated in the following
sections  represent the first plots of photon trajectories.

The plan of the paper is as follows. In Section II we study
the trajectories in Young's
double-slit experiment. In section III, we compute the trajectories corresponding to
 two down-converted photons passing through
a double-slit. In Section IV we plot the Bohmian trajectories for reflection
and refraction through a glass slab. We make some concluding remarks in Section V.

\section{Single photon double-slit interference}

Let us now consider the specific case of double-slit interference
of single photons. 
If the slits $A$ and $B$ have a
non-zero width $d$ significantly larger than the de Broglie
wavelength of the particles ($d>>\lambda$), the slits will convert
plane incident waves into plane diffracted waves sufficiently far
from them (the case of Fraunhoffer diffraction). One can see this
by carrying out the necessary approximations \cite{Born} on the
single-particle spherical wave at a point $P$,
arriving from a point within a slit at a
distance $x = \pm \xi $ from the origin, and integrating over
the slit\cite{Ghose2}. The wave function at a point $(x,y)$ at a 
sufficient distance $D >> d^2/\lambda$ to the right of the plane
of the slits is given by\footnote{Henceforth we shall write $\psi$ in place of
$\Gamma\psi$ for brevity of notation.}

\be
\psi(x,y) = {\cal M}_A g_A {{\rm exp}(ikr_A) \over r_A} + 
{\cal M}_B g_B {{\rm exp}(ikr_B) \over r_B}  
\ee
where $g_A$ and $g_B$ are the diffraction factors given by
\be
g_{A,B}= \frac{{\rm sin} (k y d /2 D)}{k y d /2 D}
\ee
and ${\cal M}_A$ and ${\cal M}_B$ are the Kemmer-Duffin wave functions given by
\be
{\cal M}_A= \pmatrix{-E_0 {\rm sin}(\theta_A) \cr -E_0 {\rm cos}(\theta_A) \cr 0 \cr 0 \cr 0
\cr B_0 \cr 0 \cr 0 \cr 0 \cr 0}
\ee
and
\be
{\cal M}_B= \pmatrix{E_0 {\rm sin}(\theta_B) \cr -E_0 {\rm cos}(\theta_B) \cr 0 \cr 0 \cr 0
\cr B_0 \cr 0 \cr 0 \cr 0 \cr 0}
\ee
where $\theta_A$ and $\theta_B$ are the angles of diffraction from slits $A$ and $B$
respectively.
  
Using the above wave function the components for Bohmian veloccity are given by
\ben
v_x = {2E_0B_0 \over \psi^{\dagger} \psi} \Biggl(g_A^2 {\rm cos}(\theta_A) +
g_B^2{\rm cos}(\theta_B) + g_Ag_B{\rm cos}[k(r_A-r_B)]({\rm cos}(\theta_A) + 
{\rm cos}(\theta_B))\Biggr) \cr
v_y = {2E_0B_0 \over \psi^{\dagger} \psi} \Biggl(-g_A^2 {\rm sin}(\theta_A) +
g_B^2{\rm sin}(\theta_B) + g_Ag_B{\rm cos}[k(r_A-r_B)]({\rm sin}(\theta_B) - 
{\rm sin}(\theta_A))\Biggr)
\een
with $\psi^{\dagger}\psi$ given by
\be
\psi^{\dagger}\psi = (E_0^2 + B_0^2)(g_A^2 + g_B^2) + 2g_Ag_B\biggl(E_0^2{\rm cos}(\theta_A + \theta_B) +
B_0^2\biggr){\rm cos}[k(r_A-r_B)]
\ee

The Bohmian trajectories for photons can now be plotted for different initial positions along
the slits using the above expressions for the velocity. We have
taken a uniform distribution of the initial
positions for both the slits. (See Figure 1). The  trajectories clearly correspond to
the probability density obtained from standard quantum theory at any line parallel to the
line joining the slits ($y$-axis). The trajectories  are
 similar to the 
trajectories  of massive particles\cite{Holland}.

\section{Bohmian trajectories of a pair of down-converted photons}

Before we proceed to compute the Bohmian trajectories for a pair of photons,
the following point needs to be clarified.
 Let us define the rank-2 tensor current
\be
s_{\mu\nu}(x_1,x_2) = c\bar{\psi}(x_1,x_2)\biggl(\beta^{(1)}_{\mu}\beta^{(1)}_{\lambda} +
\beta^{(1)}_{\lambda}\beta^{(1)}_{\mu} - g_{\mu\lambda}\biggr)a^{\lambda}
\biggl(\beta^{(2)}_{\nu}\beta^{(2)}_{\rho} + \beta^{(2)}_{\rho}\beta^{(2)}_{\nu} - g_{\nu\rho}\biggr)a^{\rho}\Gamma\psi(x_1,x_2)
\ee
for wave functions which satisfy the symmetry $\psi(x_1,x_2)=\psi(x_2,x_1)$.
Then the i-th component of the Bohmian velocity for the n-th particle ($n=1,2$) is
\be
v_i^{(n)}(x_1,x_2) = c{s_{i0}(x_1,x_2) \over s_{00}(x_1,x_2)}
\ee
Using similar arguments to those presented in section I, it is clear
that this Bohmian velocity is also time-like.
The expression (27) however appears to be non-covariant because the two sides transform differently.
Nevertheless, it is possible to write it in a manifestly covariant form by introducing
a foliation of spacetime with spacelike hypersurfaces $\Sigma$ with future oriented unit normals
$\eta^{\mu}(x)$ at every point $x$ of $\Sigma$ such that  $\eta^{\mu}(x)\eta_{\mu}(x) =1$. Then
\ben
v_i^{(1)}(x_1,x_2) = c{s_{i\mu}(x_1,x_2)\eta^{\mu}(x_2) \over s_{\mu\nu}(x_1,x_2)\eta^{\mu}(x_1)\eta^{\nu}(x_2)} \nonumber \\
v_i^{(2)}(x_1,x_2) = c{s_{i\mu}(x_1,x_2)\eta^{\mu}(x_1) \over s_{\mu\nu}(x_1,x_2)\eta^{\mu}(x_1)\eta^{\nu}(x_2)} 
\een
The fact that EPR entangled states can be written in a manifestly covariant form
using this technique of spacetime foliation was first shown by Ghose and Home\cite{Ghose3}. 
The same technique was used by Durr et al.\cite{Durr} and Holland\cite{Holland2} in the
context of Bohmian velocities for multiparticle entangled states to demonstrate their
relativistic covariance. For further details,
see \cite{Durr,Holland2}.

We now consider an experiment in which a pair of down converted photons
is made to pass through two identical slits. We will compute
the Bohmian trajectories for this case in the limit of Fraunhoffer diffraction.
The two-particle wave function (in the Fraunhoffer limit, i.e., $x_1=x_2=D>>d^2/\lambda$) is given by 

\be
\psi(y_1,y_2) = {{\rm exp}(2ikD) \over D^2} d^2 g_1g_2
\Biggl({\cal M}_{A1}{\cal M}_{B2}{\rm exp}(-ika(y_1-y_2)/D) +
{\cal M}_{A2}{\cal M}_{B1}{\rm exp}(ika(y_1-y_2)/D)\Biggr)
\ee
After substituting the expressions for the Kemmer-Duffin matrix elements
and the diffraction factors, we obtain the following expressions for
the Bohmian velocities of the two photons:

\be
v_{1x} = {c \over \psi^{\dagger}\psi}\Biggl(-2\biggl(g_1^4{\rm cos}(\theta_A)
+ g_2^4{\rm cos}(\theta_B)\biggr) + g_1^2g_2^2\Bigl(1+{\rm cos}(\theta_A + \theta_B)
\Bigr)\Bigl({\rm cos}(\theta_A) + {\rm cos}(\theta_B)\Bigr)\Biggr)
\ee

\be 
v_{2x} = v_{1x}
\ee

\be  
v_{1y} = {c \over \psi^{\dagger}\psi}\Biggl(-2\biggl(g_1^4{\rm sin}(\theta_A)
- g_2^4{\rm sin}(\theta_B)\biggr) + g_1^2g_2^2\Bigl(1+{\rm cos}(\theta_A + \theta_B)
\Bigr)\Bigl(-{\rm sin}(\theta_A) + {\rm sin}(\theta_B)\Bigr)\Biggr)
\ee

\be
v_{2y} = -v_{1y}
\ee
where $\psi^{\dagger}\psi$ is given by
\be
\psi^{\dagger}\psi = {8d^4g_1^2g_2^2E_0^2B_0^2 \over D^4}\biggl[1 + {({\rm cos}(\theta_A +
\theta_B)+1)^2 \over 4}{\rm cos}\bigl(2ka(y_1-y_2)/D\bigr)\biggr]
\ee
The second cosine term represents a fourth order interference in the joint detection
probability of the two photons\cite{Mandel}.

The Bohmian trajectories are plotted in Figure 2. It can be checked that they
agree with the joint detection probability amplitude obtained on a plane
parallel to the plane of the slits. Again, they are similar to the trajectories
one obtains for the symmetrized wave function 
of two massive particles\cite{Holland}. Note that the trajectories are symmetric about
the x-axis, and the trajectories in the upper and lower half-planes
do not cross.

\section{Reflection and Refraction through a glass slab}

Finally, let us consider the example of refraction of light through a glass
slab. We consider both the  air-glass interfaces separately and combine the solutions.
To obtain the solution of the Kemmer-Duffin equation in this case, we
must first solve
 Maxwell's equations for this case. Let the electric field 
be polarized along the $y$ direction and let the wave  propagate
along the $x$ direction. The air-glass interface is taken at
$x=0$. Let the amplitude of the electric field be represented by a gaussian
wave packet\footnote{Such wave packets are nowadays routinely produced in the
laboratory in down-conversion experiments. See, for example\cite{Chiao}.} centered at $x_0$, i.e.,
\ben
 E_x = E_0{\rm exp}\Biggl({-(x-ct-x_0)^2 \over 2\sigma_0}\Biggr)
\nonumber \\
E_y = 0 \nonumber \\
E_z = 0 
\een
Taking into account the boundary conditions at the air-glass interface, one
obtains
\be
E_x(x,t) = \pmatrix{\Biggl(E_0{\rm exp}\biggl({-(x-ct-x_0)^2 
\over 2\sigma_0}\biggr) + {1-n \over 1+n}\Biggl(E_0{\rm exp}\biggl({-(x+ct-x_0)^2
\over 2\sigma_0}\biggr)\Biggr), \, \, x < 0 \cr
\Biggl({2\over 1+n}E_0{\rm exp}\biggl({-(nx-ct-x_0)^2 
\over 2\sigma_0}\biggr)\Biggr), \, \,  x \ge 0}
\ee
where $n$ is the refractive index.
The corresponding magnetic field is given by 

\be
B_z(x,t) = \pmatrix{\Biggl({E_0 \over c}{\rm exp}\biggl({-(x-ct-x_0)^2 
\over 2\sigma_0}\biggr) - {1-n \over 1+n}\Biggl({E_0 \over c}{\rm exp}\biggl({-(x+ct-x_0)^2
\over 2\sigma_0}\biggr)\Biggr), \, \,  x < 0 \cr
\Biggl({2\over 1+n}{E_0 \over c}{\rm exp}\biggl({-(nx-ct-x_0)^2 
\over 2\sigma_0}\biggr)\Biggr), \, \, x \ge 0}
\ee

The Kemmer-Duffin-Harish Chandra wave function is therefore given by

\be
\psi = \pmatrix{-E_x \cr - E_y \cr -E_z \cr B_x \cr B_y \cr B_z 
\cr 0 \cr 0 \cr 0 \cr 0}
\ee

Using these expressions for the electric and magnetic fields, one obtains
the Bohmian velocity to be

\be
v_x = \pmatrix{c\Biggl({{\rm exp}\biggl({-(x-ct-x_0)^2 
\over \sigma_0}\biggr) - {(1-n)^2 \over (1+n)^2}\Biggl(E_0{\rm exp}\biggl({-(x+ct+x_0)^2
\over \sigma_0}\biggr)\Biggr) \over 
{\rm exp}\biggl({-(x-ct-x_0)^2
\over \sigma_0}\biggr) + {(1-n)^2 \over (1+n)^2}\Biggl(E_0{\rm exp}\biggl({-(x+ct-x_0)^2
\over 2\sigma_0}\biggr)\Biggr)}\Biggr), \, \, x < 0 \cr
\biggl({c \over n}\biggr),  x \ge 0}
\ee

Similarly, the solutions for the electric and magnetic fields, and
the corresponding expressions for Bohmian velocity can be obtained
for reflection and refraction at the next glass-air
interface placed at $x=0.2$. These two sets of solutions are combined to obtain the
trajectories for photons reflected and transmitted through the
glass slab.
The trjectories for a particular set of initial positions are plotted in Figure 3.

\section{Conclusions}

Bohm and his coworkers have all along emphasized a fundamental difference between
fermions and bosons in that fermions, in their view, are particles, whereas bosons are
fields. This asymmetry in the Bohmian picture of fermions and bosons arose due to the
absence, in their view, of a consistent relativistic quantum mechanics of bosons with
a conserved four-vector current which is time-like and whose time component is positive. Such a formulation 
was provided by Ghose et al.\cite{Ghose,Ghose3} and it was shown that Bohmian
trajectories for relativistic bosons could be defined\cite{Home}.
Just as the actual plotting of Bohmian trajectories for nonrelativistic particles
was an important advance\cite{Dewdney}, it is equally important to demonstrate
the actual nature of Bohmian trajectories for relativistic bosons in simple physical
situations, particularly because such trajectories were thought not to exist by Bohm
himself. This does not in any way detract from the significance of Bohm's general
point of view regarding the causal interpretation. In our view these trajectories
constitute a significant support of Bohm's causal interpretation
by removing an unnecessary asymmetry between fermions and bosons from it. In case there
is any truth in supersymmetry, such an asymmetry would be fatal for Bohmian mechanics.

\vskip 0.2in

Acknowledgements: PG acknowledges financial support from DST, Govt. of India. 
SG and JS thank the S. N. Bose National Centre for Basic Sciences
 for hospitality and financial support.

\noindent
Figure Captions:

\noindent
Figure 1. Bohmian trajectories of photons for self-interference through a pair
of identical slits centered at $y=-0.0002$ and $y=0.0002$.

\noindent
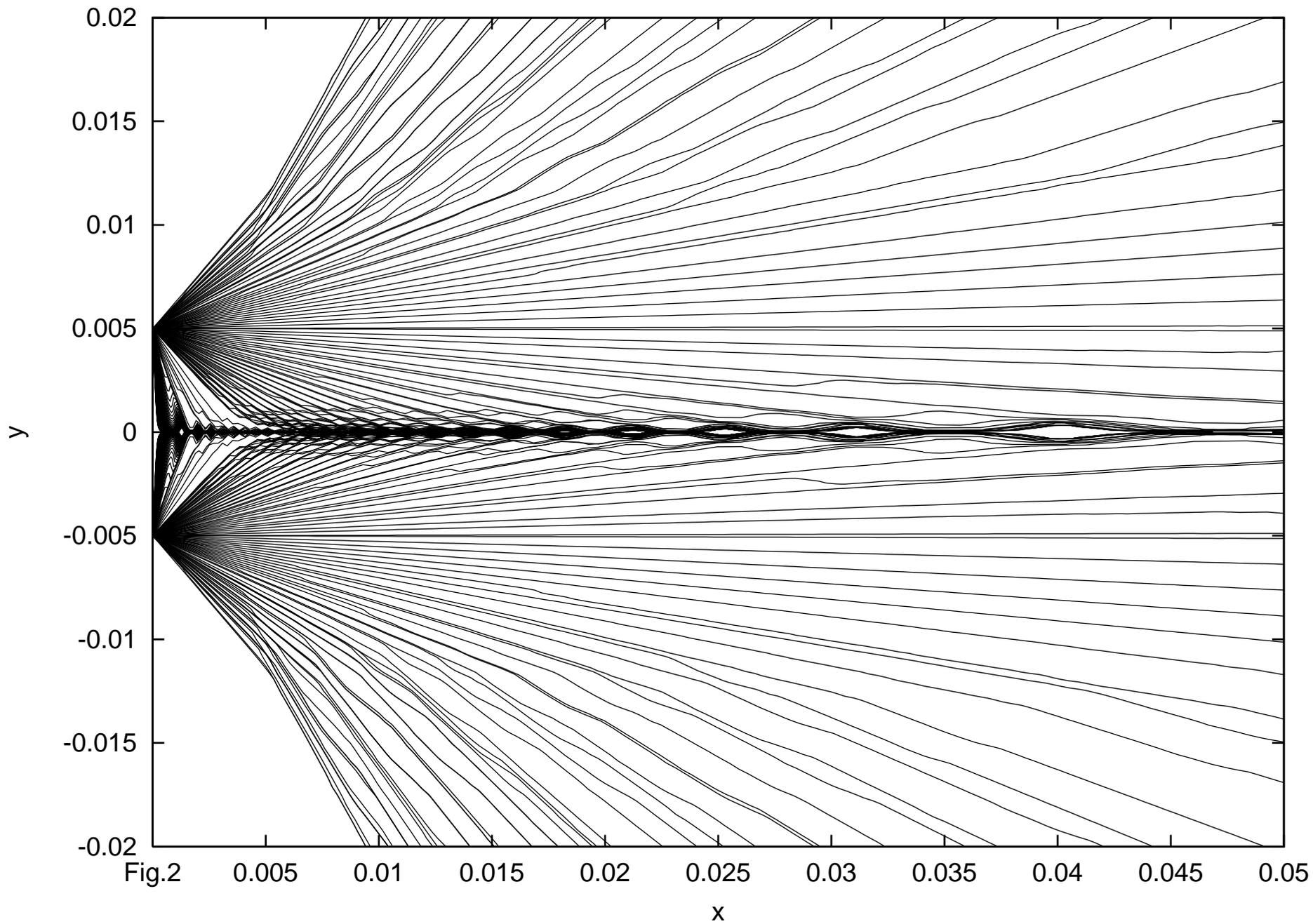
Figure 2. Bohmian trajectories for a pair of photons passing through two identical
slits. Note that there is no crossing of trajectories between the upper and lower
half planes.

\noindent
Figure 3. Bohmian trajectories for photons passing through a glass slab placed at
$0 \le x \le 0.2$. Reflection and refraction are seen at both the air-glass interfaces.


\begin{thebibliography}{999}

\bibitem{Ghose}
P. Ghose, D. Home and M. N. Sinha Roy, Phys. Lett. A{\bf 183}, 267 (1993);
P. Ghose, {\it Found. of Physics}  {\bf 26}, 1441, 1996.

\bibitem{Kemmer}
N. Kemmer, {\it Proc. Roy. Soc. A}  {\bf 173}, 91, 1939.

\bibitem{HC}
Harish-Chandra, {\it Proc. Roy. Soc. A}	 {\bf 186}, 502, 1946.

\bibitem{Prosser}
R. D. Prosser, Int. Jour. Theor. Phys. {\bf 15}, 169 (1976).

\bibitem{Born} 
M. Born and E. Wolf, {\it Principles of Optics}, Sixth Edition,
Cambridge University Press, England, 1980.

\bibitem{Ghose2}
P. Ghose, quant-ph/0001024.

\bibitem{Holland}
P. R. Holland, {\it The quantum Theroy of motion}, Cambridge University
Press, London (1993).

\bibitem{Ghose3}
P. Ghose and D. Home, Phys. Rev. A{\bf 43}, 6382 (1991).

\bibitem{Durr} D. Durr, S. Goldstein, K. M-Berndl and N. Zanghi, Phys. Rev.
A{\bf 60}, 2729 (1999).

\bibitem{Holland2} P. Holland, Phys. Rev. A{\bf 60}, 4326 (1999).

\bibitem{Mandel} R. Ghosh and L. Mandel, Phys. Rev. Lett. {\bf 59}, 1903 (1987).

\bibitem{Chiao} A. M. Steinberg and R. Y. Chiao, Phys. Rev. A{\bf 49}, 3283 (1994).

\bibitem{Home} P. Ghose and D. Home, Phys. Lett. A{\bf 191}, 362 (1994).

\bibitem{Dewdney} C. Dewdney and B. J. Hiley, Found. Phys. {\bf 12}, 27 (1982);
C. Dewdney, Phys. Lett. A{\bf 109}, 377 (1985).

\end{thebibliography}
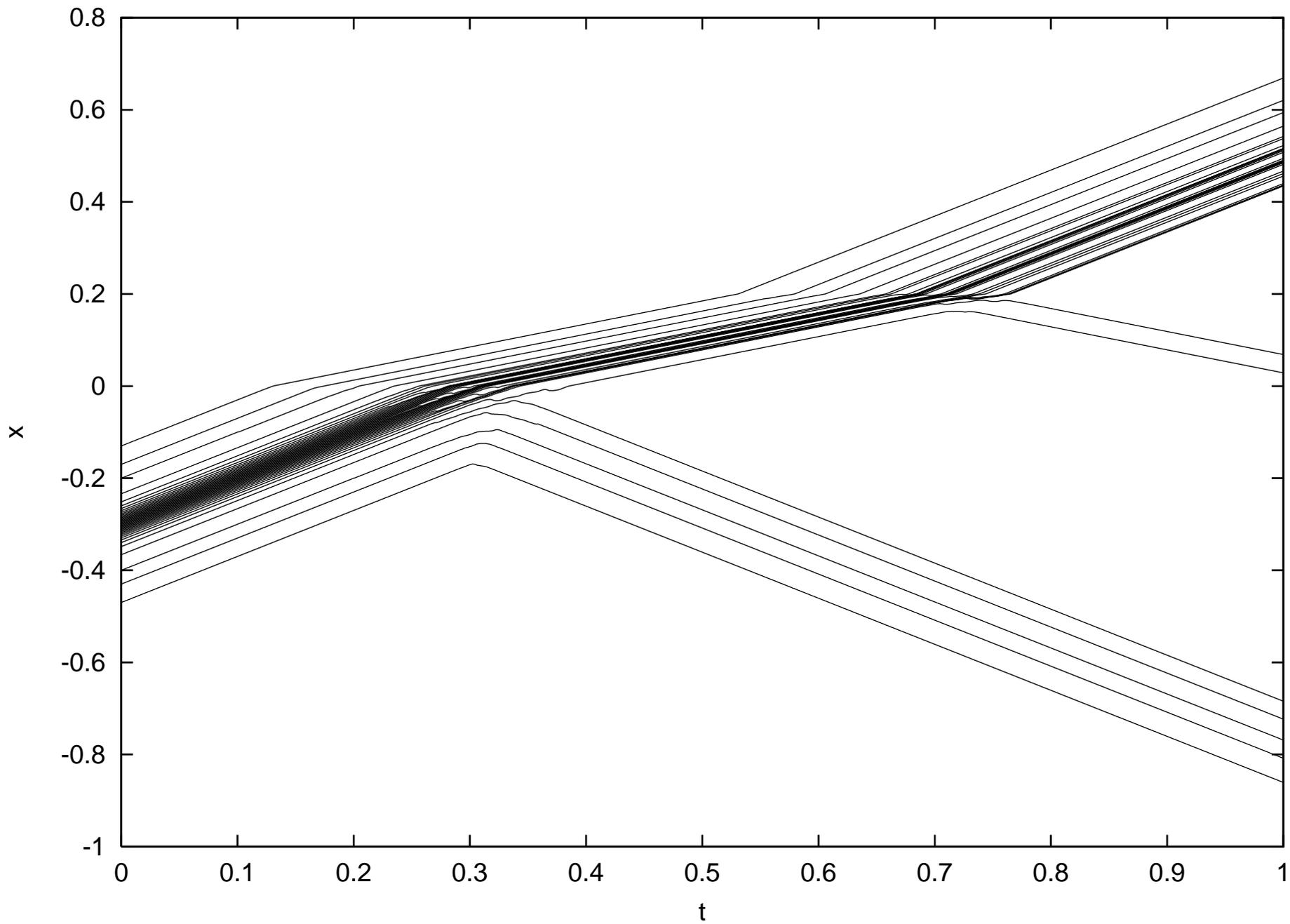
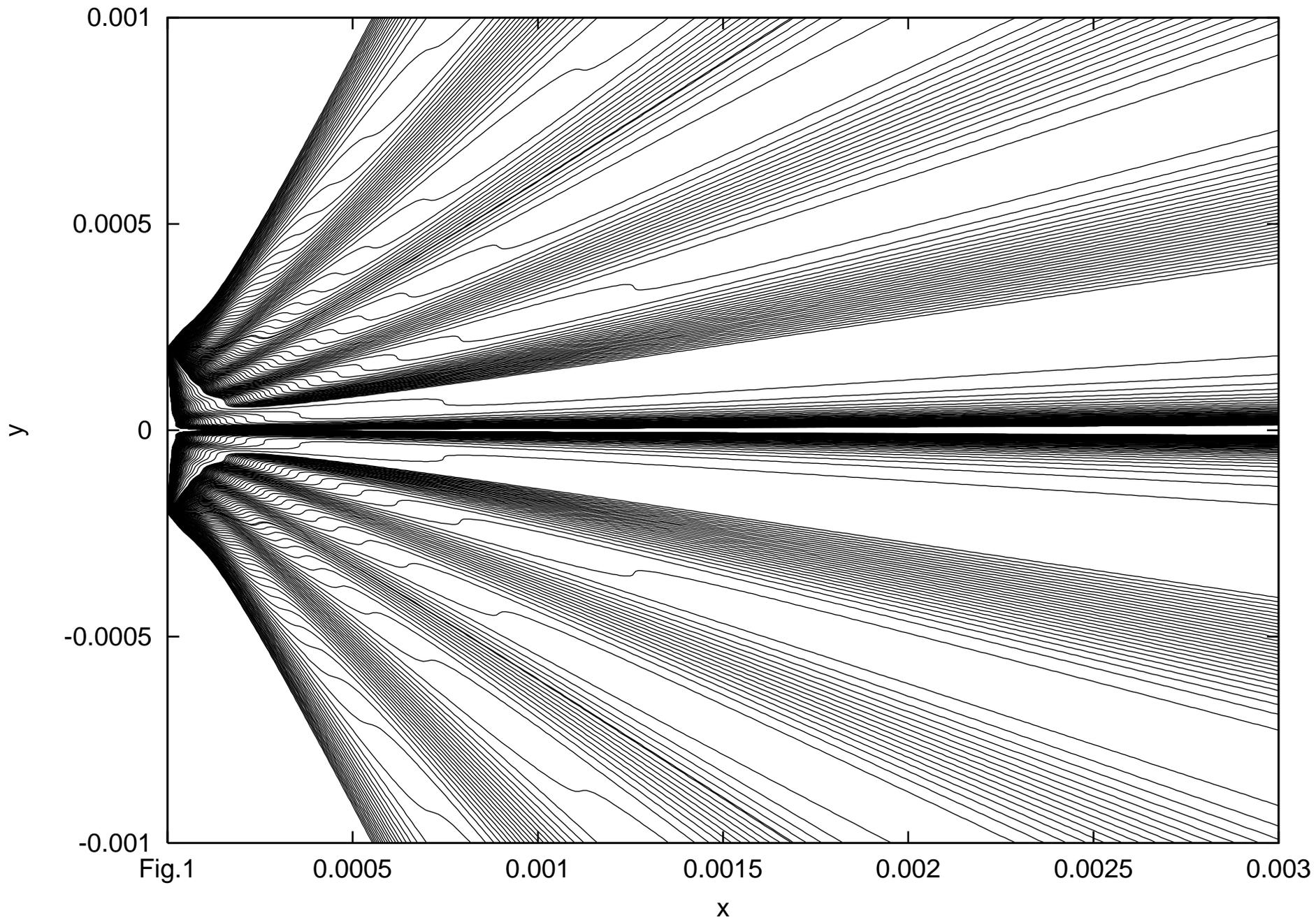
\end{document}